\documentclass[prl,preprint,floatfix]{revtex4}
\usepackage{graphicx}
\def\pla#1#2#3{Phys.  Lett.  A {\bf #1}, #2 (#3)}
\def\pre#1#2#3{Phys.  Rev.  E {\bf #1}, #2 (#3)} 
\def\prl#1#2#3{Phys.  Rev. Lett. {\bf #1}, #2 (#3)} 
\begin{document} 
\title{Robustness of spatiotemporal regularity under parametric
  fluctuations in a network of coupled chaotic maps} \title{Robust
  spatiotemporal regularity under parametric fluctuations in the
  network of coupled chaotic maps} \title{Under what kind of
  parametric fluctuations is spatiotemporal regularity the most
  robust?}  
\author{Manish Dev Shrimali{\footnote{e-mail: m.shrimali@gmail.com}}}
\affiliation{Department of Physics, Dayanand College, Ajmer, 305 001,
  India} 
\author{Swarup Poria {\footnote{e-mail:
      swarup$_-$p@yahoo.com}}} \affiliation{Department of Mathematics,
  Midnapore College, Midnapore, 721 101, West Bengal, India}
\author{Sudeshna Sinha{\footnote{e-mail:
      sudeshna@imsc.res.in}}} \address{The Institute of Mathematical
  Sciences, Taramani, Chennai 600 113, India}

\begin{abstract}
  It was observed that the spatiotemporal chaos in lattices of coupled
  chaotic maps was suppressed to a spatiotemporal fixed point when
  some fraction of the regular coupling connections were replaced by
  random links. Here we investigate the effects of different kinds of
  parametric fluctuations on the robustness of this spatiotemporal
  fixed point regime. In particular we study the spatiotemporal
  dynamics of the network with noisy interaction parameters, namely
  fluctuating fraction of random links and fluctuating coupling
  strengths. We consider three types of fluctuations: (i) noisy in
  time, but homogeneous in space; (ii) noisy in space, but fixed in
  time; (iii) noisy in both space and time. We find that the effect of
  different kinds of parameteric noise on the dynamics is quite
  distinct: quenched spatial fluctuations are the most detrimental to
  spatiotemporal regularity; spatiotemporal fluctuations yield
  phenomena similar to that observed when parameters are held constant
  at the mean-value; and interestingly, spatiotemporal regularity is
  most robust under spatially uniform temporal fluctuations, which in
  fact yields a {\em larger} fixed point range than that obtained
  under constant mean-value parameters.

\end{abstract}
\pacs{89.75Hc,05.45.-a}

\maketitle


\section{Introduction}

One of the important prototypes of extended complex systems are
nonlinear dynamical systems with spatially distributed degrees of
freedom, or alternately spatial systems composed of large numbers of
low dimensional nonlinear systems. The basic ingredients of such
systems are: (i) creation of local chaos or local instability by a low
dimensional mechanism and (ii) spatial transmission of energy and
information by coupling connections of varying strengths and
underlying topologies.

The Coupled Map Lattice (CML) is such a model, capturing the essential
features of the nonlinear dynamics of extended systems
\cite{kaneko}. A very well-studied coupling form in CMLs is nearest
neighbour coupling.  However some degree of randomness in spatial
coupling can be closer to physical reality than strict nearest
neighbour scenarios. In fact many systems of biological, technological
and physical significance are better described by randomising some
fraction of the regular links \cite{RMP_bar, Watts}. Here we focus on
a ring of coupled chaotic maps whose coupling connections are
dynamically rewired to random sites with probability $p$, namely at
any instance of time, with probability $p$ a regular link is switched
to a random one \cite{dyn_net}.

It has recently been found that such random coupling yields a
spatiotemporal fixed point in a network of chaotic maps
\cite{ss-pre02}. That is, the strongly unstable fixed point of the
local chaotic map is stabilized under increasing randomness in the
coupling connections. Thus interestingly, the inherent chaos present
in the individual local units is suppressed by dynamically switched
random links, giving rise to a global spatiotemporal fixed point
attractor.

In this paper we study the effect of {\em parametric fluctuations} on
the synchronization properties of such networks. We consider different
types of noise in the parameters: (i) spatial (ii) temporal and (ii)
both spatial and temporal. Keeping the local dynamics always fully
chaotic, we focus on parameteric noise in the interaction parameters
between the nodes in the network. In particular, we consider
fluctuations in the fraction of random links $p$ in the system, and
fluctuations in the coupling strength of the different links. That is,
we study perturbations in both the geometry of the network
connections, as reflected in noisy $p$, as well as in the strength of
the links.

\section{Model}

We consider a network of N coupled logistic maps. The sites are
denoted by integers $i = 1, \dots, N$, where $N$ is the size of the
lattice. On each site is defined a continuous state variable denoted
by $x_n (i)$, which corresponds to the physical variable of interest.

The evolution of this lattice, under interactions with the nearest
neighbours, is given by 
\begin{equation}
x_{n+1} (i) = (1 - \epsilon) f(x_n (i)) + \frac{\epsilon}{2} \{ x_n (i+1)+x_n(i-1) \}
\end{equation}
The strength of coupling is given by $\epsilon$. The local on-site map
is chosen to be the fully chaotic logistic map: $f(x) = \alpha x (1 -
x)$ with $\alpha=4$, as this map has widespread relevance as a prototype
of low dimensional chaos.

We consider the above system with its coupling connections rewired
randomly with probability $p$. Namely, at every update we will connect
a site with probability $p$ to randomly chosen sites, and with
probability $(1-p)$ to nearest neighbours, as in Eqn.~1.  That is, at
every instant a fraction $p$ of randomly chosen nearest neighbour
links are replaced by random links. The case of $p = 0$ corresponds to
the usual nearest neighbour interaction, while $p = 1$, corresponds to
completely random coupling. This type of connectivity has been
observed in a range of natural and human-engineered systems
\cite{Watts}.

In this work, we will focus on the effect of fluctuations in the
interaction parameters, i.e. noisy coupling strength $\epsilon$, and
fraction of random links $p$ \cite{ss}. Now, one can have four
distinct scenarios (denoting the relevant parameter as $A$):

(i) $A_{n}(i)\equiv A_0$

Here the parameter is constant. We will denote this case of {\em zero
  fluctuations} by $C$.

(ii) $A_{n} (i) =A_0 \pm \delta A~ \eta^{i} \equiv A(i)$

Here $\delta A$ is the strength of the fluctuation in the parameter
around mean value $A_0$ and $\eta^{i}$ is a zero-mean random number.  So
here the parameters are random in space but remain frozen in time,
i.e. the parameters are spatially fluctuating but temporally
invariant. In this case then, we have {\em quenched disorder} or a
basic inhomogeneity of the network links.  We will denote this case of
{\em spatial parametric fluctuations} by $S$.

(iii) $A_{n}(i)=A_0 \pm \delta A~ \eta_n \equiv A_n$.

Again $\delta A$ is the strength of the fluctuation in the parameter
around mean value $A_0$ and $\eta_n$ is a zero-mean random number. So
the fluctuation is a function of time but is site independent,
i.e. the noise in the parameter is synchronous for all the elements,
namely parameter $A$ is spatially uniform, though random in time. This
kind of a situation may arise when the system is quite uniform
intrinsically, but is subject to a {\em common} perturbation, for
instance from a common environmental influence, like say fluctuations
in the ambient temperature.  We will denote this case of {\em temporal
parametric fluctuations} by $T$.

(iv) $A_{n}(i)=A_0 \pm \delta A~ \eta_{n}^{i}$ 

Here the fluctuations are a function of both time and space. Such a
scenario describes a situation where the system is both inhomogeneous
in space and noisy in time. We will denote this case of {\em
  spatiotemporal parametric fluctations} by $ST$.

Here we consider $\eta$ to be uniformly distributed in the interval
$[0,1]$. We have simulated the system, for all the above cases, with
fluctuations in coupling strength (i.e. $A \equiv \epsilon$) and in
the fraction of random links (i.e. $A \equiv p$). The initial
conditions of the individual elements were randomly chosen in the
interval $[0,1]$, and sufficient transients were removed before
looking at the spatiotemporal profile of the network.

\section{Robustness of the spatiotemporal fixed point under parametric fluctuations}

Fig.~1 displays the bifurcation diagrams of the system under different
types of fluctuations in the fraction of random links $p$, around a
mean value of $p_0$. It is evident that the spatiotemporal fixed point
is quite robust under parametric fluctuations in general, as the range
of the spatiotemporal fixed point in coupling parameter space does not
reduce much under noisy $p$. It can also clearly be seen that the
bifurcation profile of the system under spatiotemporal fluctuations in
$p$ (Fig.~1d) is very similar to the system under constant $p=p_0$
(Fig.~1a). Further we observe that quenched spatial fluctuations
reduces the range of stability of the spatiotemporal fixed point most
significantly (Fig.~1b). On the other hand temporal fluctuations in
$p$ does not degrade the stability of the fixed point regime, and is
most conducive to spatiotemporal regularity. In fact, interestingly
the fixed point range obtained under temporal prametric fluctuations
(Fig.~1c) is {\em larger} than that obtained from the constant case
(Fig.~1a). Qualitatively similar bifurcation diagrams with respect to
$p$ were obtained under fluctuations in the coupling strength of the
different links.

In order to quantify the above observation we calculate the average
deviation of the system from a synchronized state, denoted
by $Z$, and defined as:
\begin{equation}
Z = << (x_n(i) - \overline{x})^2 >>
\end{equation}
where $\overline{x}$ is the mean value of $x$. The averages $<< \dots >>$
are over all sites $i$ ($i = 1, \dots N$) and over long times $n$.

Figs.~2-3 show error $Z$ with respect to coupling strength and the
fraction of random links $p$, under parametric fluctuations in $p$ and
$\epsilon$ respectively. It is clearly evident from both figures that
temporal parametric fluctuations ($T$) gives zero error for the
largest range. Spatiotemporal parametric fluctuations ($ST$) and the
case where the parameter is kept constant at the mean value ($C$) give
completely similar trends. As observed earlier, the case of spatial
paramteric fluctuations ($S$) gives the smallest range, namely the
spatiotemporal fixed point is least robust under quenched disorder.

\begin{figure}[ht]
\label{fig0}
\begin{center}
\includegraphics[width=0.8\linewidth]{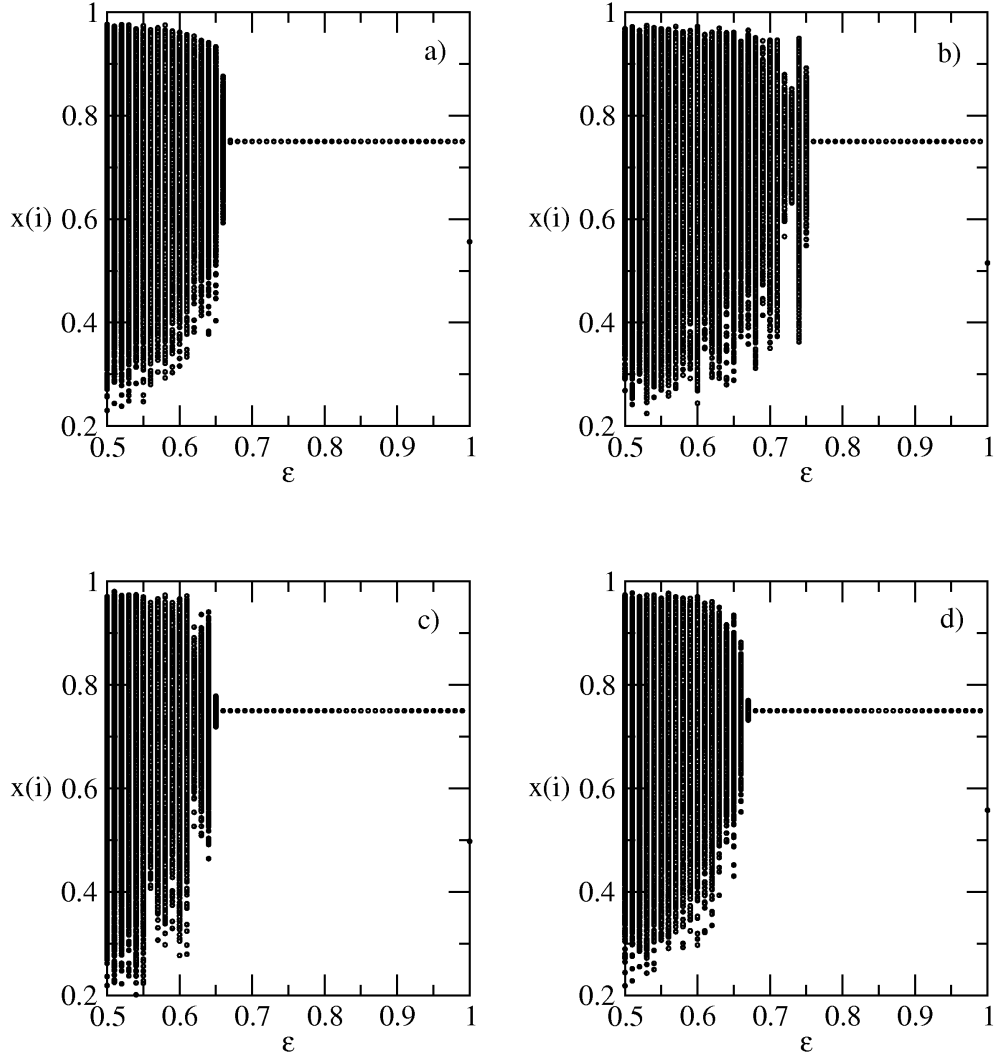}
\end{center}
\caption{State of the system with respect to coupling strength
  $\epsilon$, under different types of fluctuations in $p$: (a)
  constant $p = 0.5$ (b) spatial fluctuations (c) temporal
  fluctuations and (d) spatiotemporal fluctuations. Here strength of
  fluctuations $\delta p = 0.5$ in cases (b)-(d), around a mean value of
  $p_0=0.5$.}
\end{figure}

\begin{figure}[ht]
\label{fig1}
\begin{center}
\includegraphics[width=0.8\linewidth]{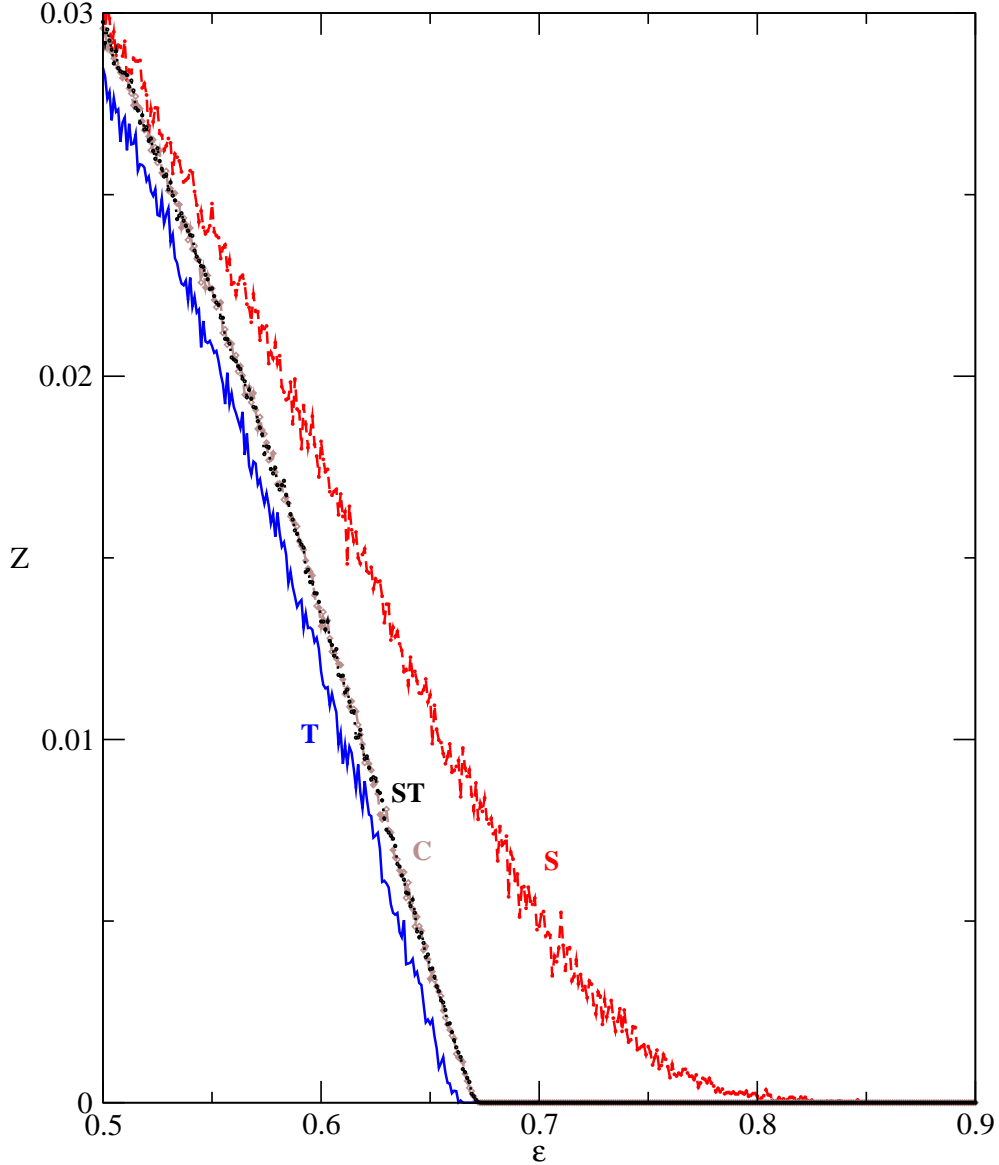}
\end{center}
\caption{Average synchronization error $Z$, as a function of coupling
  strength, for four different cases: (i) $p$ constant at the mean
  value $p_0=0.5$ denoted by $C$; (ii) spatial fluctuations in $p$
  denoted by $S$; (iii) temporal fluctuations in $p$ denoted by $T$,
  and (iv) spatiotemporal fluctuations in $p$ denoted by $ST$.  Here
  the strength of fluctuations $\delta p = 0.5$ in cases (ii)-(iv),
  namely $p$ is distributed uniformly in the range $[0 : 1]$. Observe
  that $T$ gives (almost) zero error for the largest range, $ST$ and
  $C$ give similar trends, while $S$ gives the smallest range, namely
  the least robustness for the spatiotemporal fixed point.}
\end{figure}

\begin{figure}[ht]
\label{fig2}
\begin{center}
\includegraphics[width=0.8\linewidth]{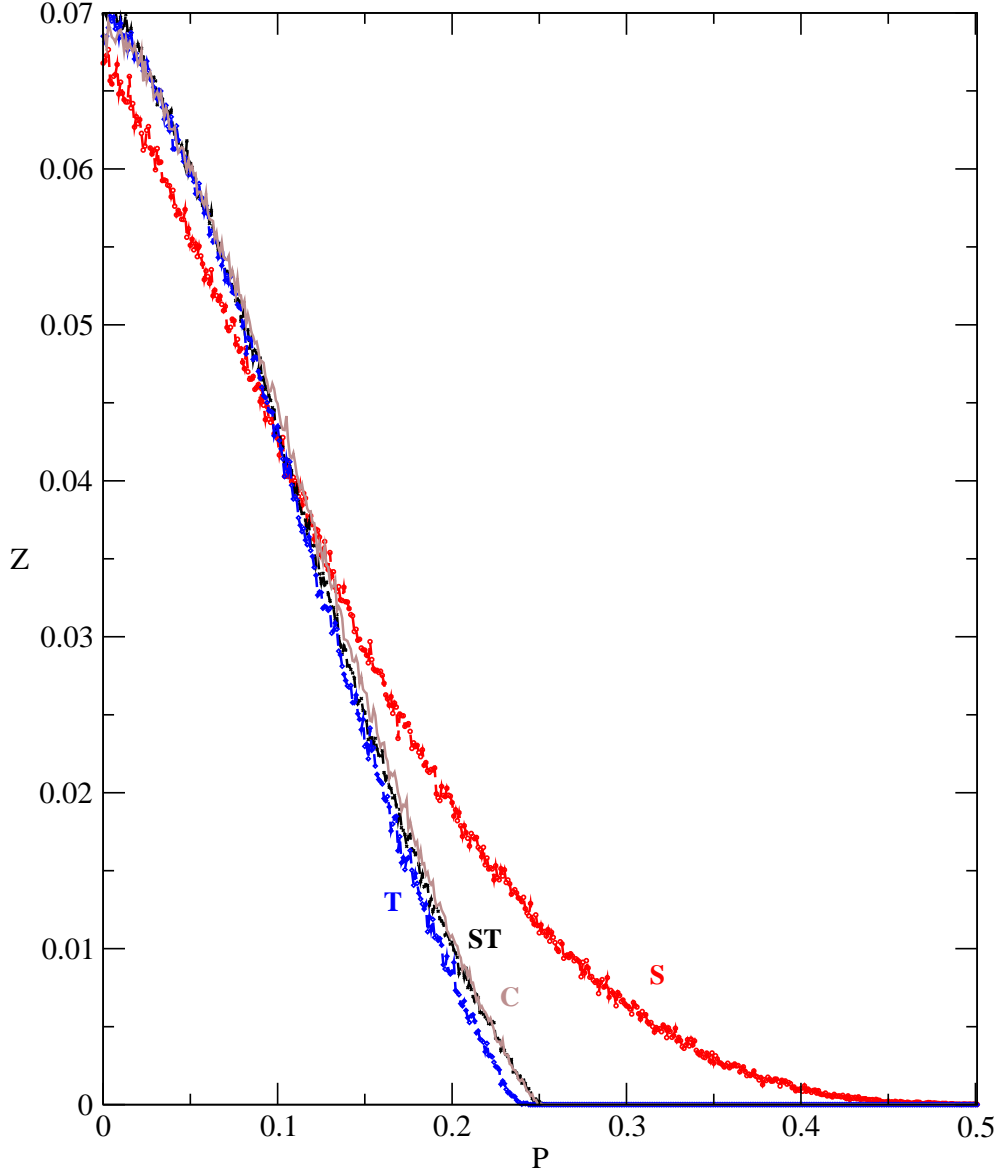}
\end{center}
\caption{Average deviation of the system from the spatiotemporal fixed
  point, $Z$, as a function of the fraction of random links $p$, for
  four different cases: (i) $\epsilon$ constant at the mean value
  $\epsilon_0 = 0.8$ denoted by $C$; (ii) spatial fluctuations in
  $\epsilon$ denoted by $S$; (iii) temporal fluctuations in $\epsilon$
  denoted by $T$, and (iv) spatiotemporal fluctuations in $\epsilon$
  denoted by $ST$. Here the strength of fluctuations $\delta \epsilon
  = 0.2$ in cases (ii)-(iv), namely $\epsilon$ is distributed
  uniformly in the range $[0.6 : 1]$. Observe that $T$ gives (almost)
  zero error for the largest range, $ST$ and $C$ give similar trends,
  while $S$ gives the smallest range, namely the least robustness for
  the spatiotemporal fixed point.}
\end{figure}

\begin{figure}[ht]
\label{fig0}
\begin{center}
\includegraphics[width=0.8\linewidth]{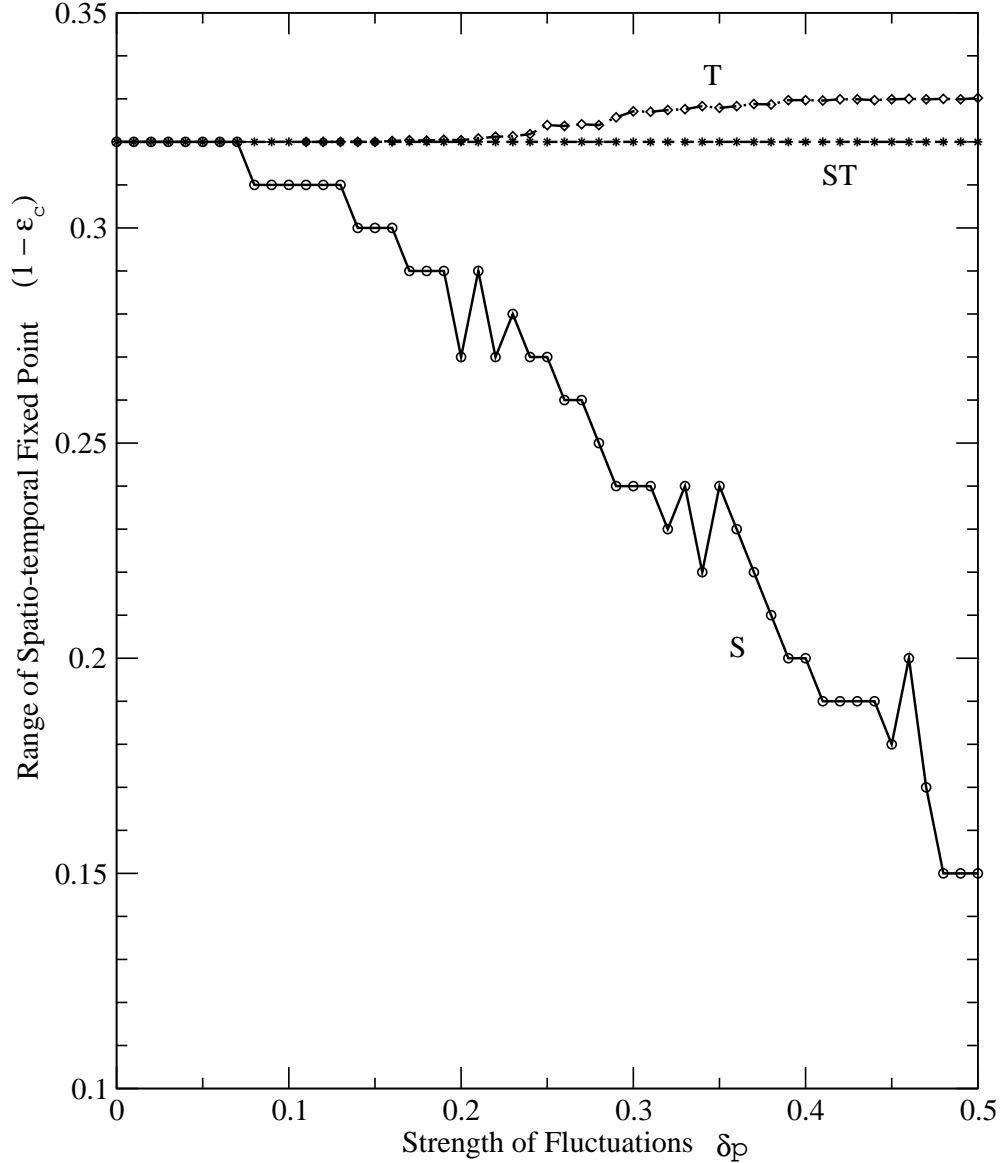}
\end{center}
\caption{Range of the spatiotemporal fixed point $(1-\epsilon_c)$,
  where $\epsilon_c$ is the critical value of coupling strength after
  which the spatiotemporal fixed point gains stability, as a function
  of the parametric fluctuation strength $\delta p$.  Note that
  fluctuation strength of $\delta p$ implies that the fraction of
  random links is distributed uniformly in the range $[p_0 - \delta p
    : p_0 + \delta p]$, with the mean $p_0 = 0.5$. Observe that
  temporal fluctuations of reasonbly large strength $\delta p$,
  actually {\em increases} the fixed point range, vis-a-vis the case
  of zero fluctuations $\delta p = 0$ (i.e. the constant $p_0$ case).}
\end{figure}

\begin{figure}[ht]
\label{fig0}
\begin{center}
\includegraphics[width=0.8\linewidth]{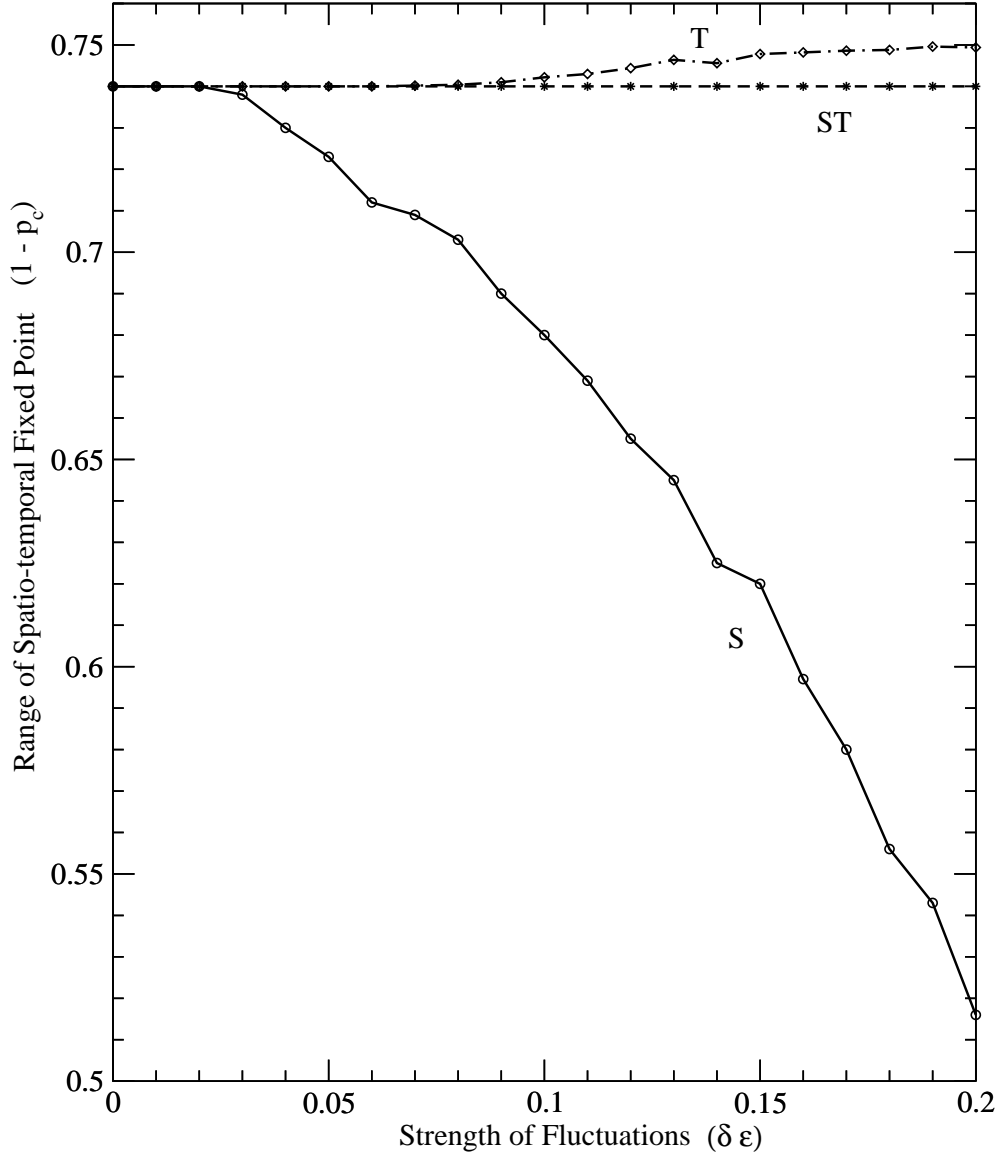}
\end{center}
\caption{Range of the spatiotemporal fixed point $(1-p_c)$, where
  $p_c$ is the critical value of $p$ after which the spatiotemporal
  fixed point gains stability, as a function of the parametric
  fluctuation strength $\delta \epsilon$.  Note that fluctuation
  strength of $\delta \epsilon$ implies that coupling $\epsilon$ is
  distributed uniformly in the range $[\epsilon_0- \delta \epsilon :
    \epsilon_0 + \delta \epsilon]$, with the mean $\epsilon_0 =
  0.8$. Observe that temporal fluctuations of reasonbly large strength
  $\delta \epsilon$, actually {\em increases} the fixed point range,
  vis-a-vis the case of zero fluctuations $\delta \epsilon = 0$
  (i.e. the constant $\epsilon_0$ case).}
\end{figure}

We also calculate the critical value of coupling strength after which
the spatiotemporal fixed point gains stability, denoted by
$\epsilon_c$, and the critical fraction of random links after which
the spatiotemporal fixed point is stable, denoted by $p_c$. The fixed
point range in coupling parameter space is then $(1 - \epsilon_c)$,
and in the space of $p$ it is $(1 - p_c)$.

Figs.~4-5, shows the range of the spatiotemporal fixed point in
$\epsilon$ and $p$ space, for varying strengths of parameteric
fluctuations. It is clear that the effect of spatiotemporal
fluctuations is quite indistinguishable from the mean field case
(namely the case of $p_0$ and $\epsilon_0$, with $\delta p$ and
$\delta \epsilon$ equal to zero). It is also evident that quenched
disorder reduces the stable range the most, while the fixed point
range is the largest for spatially uniform temporal fluctuations. In
fact remarkably, for temporal fluctuations, the fixed point range for
large fluctuation strengths is {\em larger} than that obtained for
zero fluctuations. These results re-inforce the conclusions drawn
from the calculations of the error function $Z$.

Lastly note that similar trends are also observed when the parametric
fluctuations are {\em periodic}, not random. For instance consider the
case where the fraction of random links $p$ is distributed in a period
$2$ cycle : $p_1$, $p_2$. For the case of quenched spatial periodic
fluctuations (denoting the value of $p$ at site $i$ as $p(i)$) we
have: $p_1 (1), p_2(2), p_1(3), p_2(4) \dots$ for all time. For
temporal periodic fluctuations we have $p(i) = p_1$ for all sites $i$
at time $n$, followed by $p(i) = p_2$ for all sites at time $n+1$, back
to $p(i) = p_1$ for all sites at time $n+2$ etc... For spatiotemporal
periodic fluctuations we have the parameters varying as a $2$-cycle in
space and time, namely $p_1(1), p_2(2), p_1(3), p_2(4) \dots$ at time
$n$ followed by $p_2(1), p_1(2), p_2(3), p_1(4) \dots$ at time $n+1$
etc... For all these cases one obtains the same qualitative behavior
as the random fluctuation cases, namely quenched spatial periodic
fluctuations are the most detrimental to spatiotemporal regularity;
space-time periodic fluctuations yield phenomena similar to that with
constant mean-values $p_0 = (p_1 + p_2)/2$; and spatiotemporal
regularity is most robust under spatially uniform periodic
fluctuations.

\section{Conclusions}

It was observed that lattices of coupled chaotic maps, with coupling
connections dynamically rewired to random sites with probability $p >
0$, gave rise to a window of spatiotemporal fixed points in coupling
parameter space. Here we investigate the effects of different kinds of
parametric fluctuations on the robustness of this spatiotemporal fixed
point regime. In particular we study the spatiotemporal dynamics of
the network with fluctuating rewiring probabilities and coupling
strengths, with the fluctuations being (a) noisy in time, homogeneous
in space, as applicable for intrinsically homogenoeus systems under
common environmental noise; (ii) noisy in space, and fixed in time,
namely quenched disorder; and (iii) noisy in both space and time..

We find that static spatial inhomogeniety, namely quenched disorder,
degrades spatiotemporal regularity most significantly. Spatiotemporal
fluctuations yield dynamical properties almost identical to networks
with the parameters held constant at the mean values. Interestingly,
spatiotemporal regularity is most robust under spatially uniform
temporal fluctuations. Such space-invariant temporal parametric noise
actually yields a regular range that is larger than that obtained for
systems with the parameters held constant at mean-value.

So the effect of different kinds of parameteric noise on
spatiotemporal regularity is quite distinct: quenched spatial
fluctuations are the most detrimental to spatiotemporal regularity;
spatiotemporal fluctuations yield phenomena similar to that with
constant mean-values; and spatiotemporal regularity is most robust
under spatially uniform temporal fluctuations.

\end{document}